\documentclass[aps,prl,reprint,superscriptaddress,amsmath,amssymb]{revtex4-2}

\usepackage{graphicx}
\usepackage{dcolumn}
\usepackage{bm}
\usepackage{color}
\usepackage{hyperref}
\usepackage{braket}

\hypersetup{
    pdftitle = {A Geometric Theory of Quantum Entanglement},
    colorlinks = true,
    citecolor = black,
    linkcolor=black,
    urlcolor=black
}
\begin{document}

\title{Geometric Theory of Quantum Entanglement}

\author{Lorenzo Capra}
\affiliation{DSFTA, University of Siena, Via Roma 56, 53100 Siena, Italy}
\affiliation{INFN Sezione di Perugia, I-06123 Perugia, Italy}
\author{Lucio de Simone}
\affiliation{DSFTA, University of Siena, Via Roma 56, 53100 Siena, Italy}
\affiliation{INFN Sezione di Perugia, I-06123 Perugia, Italy}
\author{Roberto Franzosi}
\affiliation{DSFTA, University of Siena, Via Roma 56, 53100 Siena, Italy}
\affiliation{INFN Sezione di Perugia, I-06123 Perugia, Italy}

\date{\today}

\begin{abstract}
Entanglement Distance (ED) was originally proposed as a geometric measure of entanglement derived from the Fubini-Study metric on the projective Hilbert space. Independently, the Meyer-Wallach and Scott measures quantify multipartite entanglement via linear entropy. In this work, we demonstrate that these two seemingly distinct frameworks are mathematically identical for pure states of arbitrary finite dimensions. We prove that ED arises naturally as the trace of the Fubini-Study metric tensor over the local subalgebra of observables. Crucially, this geometric unification yields a direct operational interpretation: the global entanglement of a pure state is exactly proportional to the total Quantum Fisher Information (QFI) available for local unitary estimation. This bridges abstract information geometry with quantum metrology, demonstrating that ED dynamically quantifies resourcefulness for distributed quantum sensing, identifying Heisenberg-limited sensitivity in regimes where standard variance-based witnesses fail.
\end{abstract}

\maketitle

{\it I. Introduction ---}
Differential geometric approaches to theoretical physics, and to quantum physics in particular, have led to profound advances in our understanding of nature. Notable examples include the Aharonov–Bohm effect \cite{PhysRev.115.485}, the Quantum Hall effect \cite{PhysRevLett.49.405}, the geometric (Berry) phase \cite{berry} and geometric quantum computation \cite{ZANARDI199994}.
A natural question is whether the geometric approach can lead to significant advances in quantum information, particularly in the quantification of multipartite entanglement.

Following seminal contributions to the geometric formulation of quantum mechanics by Kibble \cite{Kibble1979}, Gibbons \cite{gibbons}, and Brody and Hughston \cite{Brody2001}, it is now well established that the manifold of quantum states—the projective Hilbert space—is endowed with a natural Riemannian metric, namely the Fubini–Study metric. This metric, in particular, quantifies the distinguishability between quantum states. In this context, the main question addressed in this work is whether quantum entanglement can be characterized and, more specifically, quantified in terms of the sole geometric structure of quantum states. A second question concerns the connection between such a geometric characterization of entanglement and alternative approaches, for instance those based on entropic measures or on quantum metrology.

Indeed, the quantification of multipartite entanglement remains one of the central open problems in quantum information theory \cite{Horodecki2009,Plenio2007}, particularly in view of its fundamental role in applications such as quantum computation and quantum cryptography. While bipartite entanglement admits a well-established characterization through the von Neumann entropy of the reduced density matrix, the multipartite scenario is considerably more intricate. In this context, entanglement measures such as the Meyer--Wallach (MW) measure \cite{Meyer2002}, together with its generalization introduced by Scott \cite{Scott2004}, are formulated in terms of the average linear entropy of subsystems. Although these approaches provide operationally meaningful quantifiers, they are primarily motivated by algebraic and information-theoretic considerations, and traditionally lack a direct geometric interpretation.
%Simultaneously, the quantification of multipartite entanglement remains a central challenge in quantum information \cite{Horodecki2009,Plenio2007}. While bipartite entanglement is well-characterized by the von Neumann entropy of the reduced density matrix, multipartite measures such as the Meyer-Wallach (MW) measure \cite{Meyer2002} and its generalizations by Scott \cite{Scott2004} rely on the average linear entropy of subsystems. These measures are often justified algebraically but have traditionally lacked a foundational geometric motivation. 
Within the framework of Quantum Metrology, the Quantum Fisher Information (QFI) plays a central role in the characterization and detection of Quantum Entanglement. Nevertheless, the QFI captures only the portion of entanglement that contributes to enhanced sensitivity under parameter transformations generated by a specific observable H. Consequently, although the QFI is widely recognized as a powerful entanglement witness and as an operational quantifier of the metrological usefulness of entanglement, it cannot be regarded as a complete or universal measure of quantum entanglement.

Recently, a geometric measure known as Entanglement Distance (ED) was proposed \cite{Cocchiarella2020, Vesperini2023, Vesperini2024,DeSimone2026}, defined by the distance between a state and its infinitesimal local unitary displacement, approach that has already proved useful in the study of quantum graphs \cite{Vesperini2024b,DeSimone2025,Gnatenko21,GnatenkoL22,Gnatenko23,Gnatenko24,Gnatenko26, Gori2024}. 

In this work, we develop a unified geometric framework from which the relevant quantities employed in both entropic approaches to entanglement estimation and Quantum Metrology naturally emerge. In particular, we prove that, for arbitrary finite-dimensional multipartite pure states, the geometric Entanglement Distance (ED) is mathematically isomorphic to the sum of local linear entropies, thereby providing a rigorous geometric foundation for the Meyer--Wallach family of entanglement measures \cite{Meyer2002, Scott2004}.

More importantly, we demonstrate that the ED is exactly proportional to the sum of the Quantum Fisher Information (QFI) associated with a complete set of local generators \cite{Toth2012,Toth2014,Hyllus2012}. This result establishes that, for pure states, entanglement fundamentally quantifies the metrological capability of a quantum system to detect local unitary rotations.

{\it II. Geometric and Entropic Unification ---}
The Entanglement Distance (ED) \cite{Cocchiarella2020,Vesperini2023,Vesperini2024,Vesperini2024b,DeSimone2025,DeSimone2026} is formulated within a differential geometric framework, in which a general finite-dimensional multipartite hybrid Hilbert space $\mathcal{H}$ is described through the associated projective Hilbert space $P(\mathcal{H})$, endowed with the Fubini–Study (FS) metric \cite{Brody2001, gibbons}, which defines the distance between quantum states. Within this geometric setting, quantum entanglement emerges as an intrinsic property of the local geometric structure induced by local unitary transformations.

For a multipartite quantum state  $\ket{\psi} \in \mathcal{H}=\otimes_{\mu=0}^{M-1}\mathcal{H}_\mu$, with $\dim(\mathcal{H}_\mu)=d_\mu$, local unitary (LU) transformations 
$U=\otimes_{\mu=0}^{M-1} U_\mu$, which are tensor product of local  unitary operators $U_\mu$, generate equivalence classes in the projective Hilbert space $\mathbb P(\mathcal H)$, whose elements  $\ket{U,\psi} $, share the same entanglement properties.
Considering infinitesimal LU transformations generated by the operators $dU \equiv \exp(-i \sum_\mu \sum_k  G^{(\mu)}_k d\xi^k_{\mu})$, where the $G_k^{(\mu)}$ are the generators of $su(d_\mu)$, the Fubini-Study metric 
$d^2_{FS}(\ket{U,\psi} +|dU,\psi\rangle,\ket{U,\psi})=\langle dU,\psi |dU,\psi \rangle -\langle U, \psi |dU,\psi \rangle \langle dU,\psi \ket{U,\psi}$  induces a local geometric structure on each LU orbit.
Although the corresponding metric tensor $g(\ket{U,\psi})$ depends on the orbit point, its trace is LU invariant, namely $tr[g(\ket{U,\psi})]=tr[g(\ket{\psi})]$. The Entanglement Distance is then defined as 
\begin{equation}
E(\ket{\psi}) = tr [g(\ket{\psi})]  - tr [g(\ket{\psi_\mu}^{\otimes \mu})]  \, ,
\end{equation}
where $\ket{\psi_\mu}^{\otimes \mu}$  is a fully separable state.
A direct analytical calculation leads to the explicit expression for the Entanglement Distance \cite{Cocchiarella2020}
\begin{equation}
    E(|\psi\rangle) \equiv \sum_{\mu} \frac{2(d_{\mu}-1)}{d_{\mu}} - \sum_{i} \langle \psi | G_i^{(\mu)} | \psi \rangle^2 \, .
    \label{eq:ED_def}
\end{equation}

By expanding the expectation values of the local generators $\{G^{(\mu)}_k\}_k$ in terms of the reduced density matrix associated with the $\mu$-th Hilbert subspace, $\rho_\mu = tr_{\bar{\mu}\neq \mu}[\ket{\psi}\bra{\psi}]$, Eq.~\eqref{eq:ED_def} can be exactly rewritten as
\begin{equation}
    E(|\psi\rangle) = 2 \sum_{\mu} \left( 1 - tr[\rho_\mu^2] \right) \, .
    \label{eq:ED_MW}
\end{equation}
\emph{This result demonstrates that the ED belongs, up to a normalization factor, to the Meyer–Wallach family of entanglement measures \cite{Brennen2003}. As a consequence, ED inherits LOCC invariance directly from such measures, thereby promoting it from a local unitary invariant to a full entanglement monotone}.

On the other hand, under the same assumptions, the FS distance assumes the expression of a covariance-tensor $d_{FS}^2 = \sum_{\mu,\nu} \sum_{k,r} \text{Cov}_{\ket{U,\psi}}(G^{(\mu)}_k, G^{(\nu)}_r) d\xi^k_\mu d\xi^r_\nu$, where $\text{Cov}_{|\psi\rangle}(A,B) = \bra{ \psi } A B \ket{ \psi } - \bra{ \psi } A \ket{ \psi } \bra{ \psi } B \ket{ \psi }$.

Therefore, within this framework, the trace of the metric tensor takes the form
\begin{equation}
    tr[g] = \sum_\mu \sum_k \text{Var}_{\ket{\psi}}[G^{(\mu)}_k] \, ,
    \label{eq:ED_QFI}
\end{equation}
where $\text{Var}_{\ket{\psi}}(A)=\text{Cov}_{|\psi\rangle}(A,A)$. Remarkably, Eq. \eqref{eq:ED_QFI} admits a further suggestive interpretation. Consider the set of local Hamiltonians ${H}_\mu = \sum_i v_i^\mu G_i^{(\mu)}$, where $\Vert{\bf v}^\mu\Vert^2 = 1$. The minimization of $tr[g]$ over the set $\{{\bf v}^\mu \in \mathbb{R}^{d_\mu}| \Vert{\bf v}^\mu\Vert^2 = 1\}$ leads directly to Eq.~\eqref{eq:ED_QFI}.

\emph{Equation \eqref{eq:ED_QFI} corresponds, up to a factor of 4 and a constant term, to the total Quantum Fisher Information associated with all possible local transformations. Consequently, the ED quantifies the total metrological information accessible through local operations, thereby establishing a direct theoretical connection between quantum information geometry and quantum sensing}.

{\it III. Entanglement as a Metrological Resource ---}

{\it A. Dynamical Squeezing: The Kitagawa-Ueda Model ---}
To illustrate the practical advantages of the proposed geometric framework over standard metrological witnesses, we analyze the Kitagawa-Ueda (KU) Model \cite{Kitagawa1993}, which is widely regarded as a paradigmatic model for the generation of spin squeezing \cite{Wineland1992,Ma2011} and macroscopic quantum entanglement, and has been successfully realized in atomic ensembles and Bose–Einstein condensates \cite{Gross2010,Riedel2010}. In particular, we consider an $N$-qubit state, initially fully polarized along the $x$-direction,
$
|\psi(0)\rangle = |+x\rangle^{\otimes N},
$ evolving under the one-axis twisting Hamiltonian
\begin{equation}
{H}_{\text{KU}} = \chi J_z^2 \, ,
\end{equation}
where
$
J_z = \sum_{\mu=1}^{N} \sigma_z^{(\mu)}/2
$.
Owing to the permutational symmetry of both the initial state and the Hamiltonian, all local reduced density matrices undergo identical depolarization. By applying the entropic equivalence in Eq.~\eqref{eq:ED_MW}, the Entanglement Distance can be evaluated exactly as
\begin{equation}
E(t)
=
N\left(
1-\cos^{2N-2}(\chi t)
\right).
\end{equation}

\begin{figure}[t]
    \centering
    \includegraphics[width=.9\columnwidth]{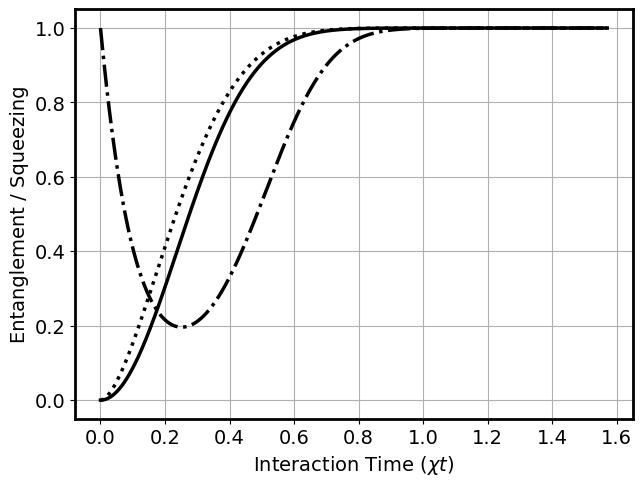} % REPLACE WITH YOUR ACTUAL IMAGE FILE
    \caption{Comparison between the standard spin squeezing parameter $\xi^2$ (dash-dotted line) \cite{Kitagawa1993}, the geometric Entanglement Distance (solid line) and the normalized Entropy of Entanglement measure for the bi-partition 1 spin/all other spins (dotted line) as a function of dimensionless time $\chi t$ for the Kitagawa-Ueda Model ($N=10$). While squeezing fails to detect entanglement after its minimum, ED monotonically tracks the formation of macroscopic non-Gaussian correlations.}
    \label{fig:kum_comparison}
\end{figure}

As shown in Fig.~\ref{fig:kum_comparison}, the standard spin-squeezing parameter $\xi^2$ successfully witnesses entanglement only during the initial Gaussian stage of the dynamics, failing once the state wraps around the Bloch sphere. In striking contrast, the ED continues to increase monotonically. At $\chi t = \pi/2$, the local Bloch vectors vanish completely, and the ED reaches its theoretical maximum, $E=N$. This geometrically signals the formation of a maximally entangled Greenberger-Horne-Zeilinger (GHZ) state--a macroscopic quantum superposition capable of achieving Heisenberg-limited sensing ($QFI \propto N^2$) in regimes where standard variance-based witnesses fail \cite{Giovannetti2004,Pezze2018}.

Crucially, our geometric result yields the exact optimal local measurement basis for extracting this metrological advantage. The optimal local direction $\vec{v}_{\text{opt}}$ aligns with the local Bloch vector $\vec{b}_\mu(t)$. Solving the Heisenberg equations of motion reveals that $\vec{b}_\mu(t) = (\cos^{N-1}(\chi t), 0, 0)^T$. This proves a highly counterintuitive and experimentally advantageous result: while the global state undergoes complex dynamical twisting, the optimal local measurement basis remains strictly static along the original axis of polarization ($x$-axis). Observers do not need to dynamically track optimal squeezing angles to access the geometric entanglement resource.

{\it B. Typicality of the Geometric Resource: Haar-Random and AME States---}
To highlight the profound difference between dynamical entanglement generation and the intrinsic geometry of the Hilbert space, we evaluate the geometric entanglement for typical pure states drawn from the Haar measure. This answers a fundamental kinematic question: what fraction of the maximum possible metrological resourcefulness does a purely random state possess, averaged across all possible macroscopic partitions?

Consider an $N$-qubit pure state drawn from the uniform Haar distribution, partitioned into a subsystem $A$ of $k$ qubits ($d_A = 2^k$) and a subsystem $B$ of $N-k$ qubits. The average purity of the reduced density matrix $\rho_A$ is governed by Lubkin's formula \cite{Lubkin1978, Popescu2006}, $\langle \text{Tr}[\rho_A^2] \rangle = (2^k + 2^{N-k})/(2^N + 1)$. 

For a specific bipartition of size $k$, the geometric entanglement is maximized when the subsystem is perfectly mixed ($\text{Tr}[\rho_A^2] = 2^{-k}$). We can therefore define a strictly normalized fractional entanglement for this cut as 
\begin{equation}
\mathcal{E}_k \equiv \frac{E_k}{E_k^{\text{max}}} = \frac{(1 - \text{Tr}[\rho_A^2])}{(1 - 2^{-k})}. 
\end{equation}

Substituting Lubkin's formula, the Haar-averaged normalized entanglement for any given partition of size $k$ simplifies exactly to $\langle \mathcal{E}_k \rangle = (2^N - 2^k)/(2^N + 1)$.

To capture the global metrological typicality of the state, we average this normalized fraction over all $2^N - 2$ proper bipartitions. Weighting by the binomial multiplicity $\binom{N}{k}$, we compute $\langle \overline{\mathcal{E}} \rangle = (2^N - 2)^{-1} \sum_{k=1}^{N-1} \binom{N}{k} \langle \mathcal{E}_k \rangle$. Evaluating the binomial sums yields a closed-form analytical expression for the global average normalized entanglement:
\begin{equation}
    \langle \overline{\mathcal{E}} \rangle_{\text{Haar}} = \frac{4^N - 3^N - 2^N + 1}{(2^N - 2)(2^N + 1)}.
    \label{eq:Haar_Normalized_Avg}
\end{equation}

In the thermodynamic limit ($N \to \infty$), the highest-order exponential terms ($4^N$) perfectly cancel, yielding:
\begin{equation}
    \lim_{N \to \infty} \langle \overline{\mathcal{E}} \rangle_{\text{Haar}} = 1.
\end{equation}

This is a profound geometric result. It rigorously proves that for a typical random state in the projective Hilbert space, the geometric entanglement approaches exactly $100\%$ of the theoretical limit across all boundaries simultaneously. Consequently, Haar-typical states natively approximate the structure of Absolutely Maximally Entangled (AME) states, possessing Heisenberg-scaled local Quantum Fisher Information across any chosen bipartition. 

However, this reveals a fundamental metrological dichotomy. While typical states theoretically saturate the geometric resource bound ($\langle \overline{\mathcal{E}} \rangle \to 1$), extracting this advantage is experimentally unfeasible without prior tomography. Because the local Bloch vectors of a Haar-random state are vanishingly small and isotropically distributed, the optimal local measurement directions $\vec{v}_{\text{opt}}$ are completely random. This stands in stark contrast to the KU model dynamics, where the symmetry of the Hamiltonian explicitly locks the optimal metrological measurement to a predictable, static axis, rendering the macroscopic quantum resource physically accessible.

This result provides a geometric analogue to Page's theorem \cite{Page1993}: the overwhelming majority of the projective Hilbert space volume is occupied by states with near-maximal Entanglement Distance, and consequently, near-maximal local Quantum Fisher Information. 

{\it IV. Conclusion ---}
In this Letter, we have established a foundational unification of three distinct paradigms in quantum information: geometry, entropy, and metrology. We demonstrated that the Entanglement Distance, defined via the Fubini-Study metric, is mathematically identical to the algebraic Meyer-Wallach measure. This geometric framework provides a direct operational interpretation: global entanglement is proportional to the total Quantum Fisher Information available through local unitary operations. 

In the Kitagawa-Ueda model, ED acts as a robust, dynamic quantifier of metrological resourcefulness, continuously tracking the formation of macroscopic GHZ-like states where standard variance-based witnesses fail. Furthermore, the geometric optimization reveals that the optimal local measurement basis is strictly static, significantly simplifying experimental protocols for extracting Heisenberg-limited sensitivity.

Finally, this unification is not merely a formal equivalence for finite-dimensional discrete systems; it provides a necessary foundation for extending entanglement measures into infinite-dimensional spaces. Traditional entropic measures often encounter severe mathematical ambiguities in Continuous Variable (CV) systems due to trace-class issues and unbounded operators. In contrast, the geometric Fubini-Study approach of the Entanglement Distance relies strictly on the variance of local generators, remaining robust even when entropic divergences occur. This paves the way for adapting ED into a universal, operationally meaningful entanglement measure for both CV and hybrid discrete-continuous systems.

\begin{acknowledgments}

The authors acknowledge the support of the Research Support Plan 2022—Call for applications for funding allocation to research projects curiosity-driven (F CUR)—Project “Entanglement Protection of Qubits’ Dynamics in a Cavity”—EPQDC and the support from the Italian National Group of Mathematical Physics (GNFM-INdAM). R.F. would like to acknowledge INFN Pisa for the financial support for this activity.
% Add any funding or grant acknowledgments here if required.
\end{acknowledgments}

{\it Appendix }

{\it A. Proof of Equivalence between Entanglement Distance and the Mayer-Wallach/Scott family of Measures.}

The state of a subsystem can be recovered from the reduced density matrix $\rho_\mu = \text{Tr}_{\bar{\mu}}[\ket{\psi}\bra{\psi}]$, where $\rho_\mu$ is a hermitian operator of unitary trace belonging to $\mathcal{L}^+(\mathcal{H}_\mu,\mathcal{H}_\mu)$. Thanks to this property $\rho_\mu$ can be decomposed in terms of the generators of the $SU(d_\mu)$ algebra $\{G^{(\mu)}_i \}$
\begin{equation}
    \rho_\mu = c^{(\mu)}_0 \mathbb{I} + \sum_{i=1}^{d_\mu^2-1} c^{(\mu)}_i G^{(\mu)}_i\,.
\end{equation}

Imposing $\text{Tr}[\rho_\mu]=1$ and remembering that $\text{Tr}[G^{(\mu)}_i]=0 \quad \forall \ i$ we get that 
\begin{equation}
    c^{(\mu)}_0 = \frac{1}{d_\mu}\,.
\end{equation}
To find the $c_i^{(\mu)}$ we study instead
\begin{equation}
    \text{Tr}[\rho_\mu G^{(\mu)}_k] = c^{(\mu)}_0 \text{Tr}[G_k] + \sum_{i=1}^{d^2_\mu-1} c^{(\mu)}_i \text{Tr}\Big[G_k^{(\mu)} G_i^{(\mu)}\Big]\,,
\end{equation}
which, exploiting the property $\text{Tr}\Big[G_k^{(\mu)} G_i^{(\mu)}\Big] = 2 \delta_{ki}$ comes out to be
\begin{equation}
    \text{Tr}[\rho_\mu G^{(\mu)}_k] = 2 c^{(\mu)}_k \qquad c^{(\mu)}_k = \frac{\langle G^{(\mu)}_k \rangle}{2}
\end{equation}

and we therefore get that
\begin{equation}\label{def::reduced_density}
    \rho_\mu = \frac{1}{d_\mu} \mathbb{I} + \sum_{i=1}^{d^2_\mu-1} \frac{\langle G^{(\mu)}_i \rangle}{2} G^{(\mu)}_i\,.
\end{equation}

Using this result we can study the Purity $\text{Tr}[\rho_\mu^2]$

\begin{multline}
        \text{Tr}[\rho_\mu^2] = \text{Tr}\Bigg[ \Big( \frac{1}{d} \mathbb{I} + \sum_{i=1}^{d^2_\mu-1} \frac{\langle G^{(\mu)}_i \rangle}{2} G^{(\mu)}_i \Big)^2 \Bigg] = \\ =\text{Tr} \Big[ \frac{1}{d^2} \mathbb{I} \Big] + \frac{1}{2} \sum_{i=1}^{d_\mu^2-1} \langle G^{(\mu)}_i \rangle^2
\end{multline}

\begin{equation}
    \text{Tr}[\rho_\mu^2] = \frac{1}{d_\mu} + \frac{1}{2} \sum_{i=1}^{d_\mu^2-1} \langle G^{(\mu)}_i \rangle^2\,,
\end{equation}

it can be noted that the sum squared of the expectations on the Generalized Gell-Mann Matrices is the same term present in Entanglement Distance, by substituting it we therefore get

\begin{equation}
    \sum_{i=1}^{d_\mu^2-1} \langle G^{(\mu)}_i \rangle^2 = 2 \Big( \text{Tr}[\rho_\mu^2]- \frac{1}{d_\mu} \Big)
\end{equation}

\begin{multline}
    E(\ket{\psi}) = \sum_\mu \frac{2(d_\mu-1)}{d_\mu} - 2 \Big( \text{Tr}[\rho_\mu^2]- \frac{1}{d_\mu} \Big) = \\ =  2\sum_\mu (1 - \text{Tr}[\rho_\mu^2])\,.
\end{multline}

We have therefore demonstrated that Entanglement Distance can be written as the sum of the linear entropies of each subsystem. The measure was already defined and used for systems of qubits by Meyer and Wallach in \cite{Meyer2002} and later explored by Brennen in \cite{Brennen2003} which showed that it could be written as 
\begin{equation}
    Q(\ket{\psi}) = 2 \Bigg( 1- \frac{1}{n} \sum_{\mu=0}^{n-1} \text{Tr}[\rho_\mu^2]\Bigg).
\end{equation}

It is to be noted that a generalization of the Meyer Wallach measure was proposed by Scott in \cite{Scott2004}, which would be a measure of entanglement for a generic system of qudits. Entanglement Distance is a special case of such formula.

\bibliographystyle{apsrev4-2} % This tells RevTeX to format it perfectly for PRL
\bibliography{bibliography} % Replace with the name of your .bib file (without the .bib extension)

@article{Kibble1979,
  title={Geometrization of Quantum Mechanics},
  author={Kibble, T.W.B.},
  journal={Communications in Mathematical Physics},
  year={1979}
}

@article{Brody2001,
   title={Geometric quantum mechanics},
   volume={38},
   ISSN={0393-0440},
   url={http://dx.doi.org/10.1016/S0393-0440(00)00052-8},
   DOI={10.1016/s0393-0440(00)00052-8},
   number={1},
   journal={Journal of Geometry and Physics},
   publisher={Elsevier BV},
   author={Brody, Dorje C. and Hughston, Lane P.},
   year={2001},
   month=apr, pages={19–53} }

@article{Meyer2002,
   title={Global entanglement in multiparticle systems},
   volume={43},
   ISSN={1089-7658},
   url={http://dx.doi.org/10.1063/1.1497700},
   DOI={10.1063/1.1497700},
   number={9},
   journal={Journal of Mathematical Physics},
   publisher={AIP Publishing},
   author={Meyer, David A. and Wallach, Nolan R.},
   year={2002},
   month=sep, pages={4273–4278} }

@article{Scott2004,
   title={Multipartite entanglement, quantum-error-correcting codes, and entangling power of quantum evolutions},
   volume={69},
   ISSN={1094-1622},
   url={http://dx.doi.org/10.1103/PhysRevA.69.052330},
   DOI={10.1103/physreva.69.052330},
   number={5},
   journal={Physical Review A},
   publisher={American Physical Society (APS)},
   author={Scott, A. J.},
   year={2004},
   month=may }

@article{Cocchiarella2020,
  title={Entanglement distance for arbitrary m-qudit hybrid systems},
  author={Cocchiarella, Denise and Scali, Stefano and Ribisi, Salvatore and Nardi, Bianca and Bel-Hadj-Aissa, Ghofrane and Franzosi, Roberto},
  journal={Physical Review A},
  volume={101},
  number={4},
  pages={042129},
  year={2020},
  publisher={APS}
}

@article{Vesperini2024,
  title={Unveiling the geometric meaning of quantum entanglement: Discrete and continuous variable systems},
  author={Vesperini, Arthur and Bel-Hadj-Aissa, Ghofrane and Capra, Lorenzo and Franzosi, Roberto},
  journal={Frontiers of Physics},
  volume={19},
  number={5},
  pages={51204},
  year={2024},
  publisher={Springer}
}

@article{Vesperini2023,
  title={Entanglement and quantum correlation measures for quantum multipartite mixed states},
  author={Vesperini, Arthur and Bel-Hadj-Aissa, Ghofrane and Franzosi, Roberto},
  journal={Scientific Reports},
  volume={13},
  number={1},
  pages={2852},
  year={2023},
  publisher={Nature Publishing Group UK London}
}

@misc{Brennen2003,
      title={An observable measure of entanglement for pure states of multi-qubit systems}, 
      author={Gavin K. Brennen},
      year={2003},
      eprint={quant-ph/0305094},
      archivePrefix={arXiv},
      primaryClass={quant-ph},
      url={https://arxiv.org/abs/quant-ph/0305094}, 
}

@article{Toth2012,
  title = {Multipartite entanglement and high-precision metrology},
  author = {T\'oth, G\'eza},
  journal = {Phys. Rev. A},
  volume = {85},
  issue = {2},
  pages = {022322},
  numpages = {8},
  year = {2012},
  month = {2},
  publisher = {American Physical Society},
  doi = {10.1103/PhysRevA.85.022322},
  url = {https://link.aps.org/doi/10.1103/PhysRevA.85.022322}
}

@article{Toth2014,
doi = {10.1088/1751-8113/47/42/424006},
url = {https://doi.org/10.1088/1751-8113/47/42/424006},
year = {2014},
month = {10},
publisher = {IOP Publishing},
volume = {47},
number = {42},
pages = {424006},
author = {Tóth, Géza and Apellaniz, Iagoba},
title = {Quantum metrology from a quantum information science perspective},
journal = {Journal of Physics A: Mathematical and Theoretical}
}

@article{Hyllus2012,
   title={Fisher information and multiparticle entanglement},
   volume={85},
   ISSN={1094-1622},
   url={http://dx.doi.org/10.1103/PhysRevA.85.022321},
   DOI={10.1103/physreva.85.022321},
   number={2},
   journal={Physical Review A},
   publisher={American Physical Society (APS)},
   author={Hyllus, Philipp and Laskowski, Wiesław and Krischek, Roland and Schwemmer, Christian and Wieczorek, Witlef and Weinfurter, Harald and Pezzé, Luca and Smerzi, Augusto},
   year={2012},
   month=feb }

@article{Kitagawa1993,
  title = {Squeezed spin states},
  author = {Kitagawa, Masahiro and Ueda, Masahito},
  journal = {Phys. Rev. A},
  volume = {47},
  issue = {6},
  pages = {5138--5143},
  numpages = {0},
  year = {1993},
  month = {6},
  publisher = {American Physical Society},
  doi = {10.1103/PhysRevA.47.5138},
  url = {https://link.aps.org/doi/10.1103/PhysRevA.47.5138}
}

@article{Lubkin1978,
  title = {Entropy of an $n$-system from its correlation with a $k$-reservoir},
  author = {Lubkin, Elihu},
  journal = {Journal of Mathematical Physics},
  volume = {19},
  number = {5},
  pages = {1028--1031},
  year = {1978},
  publisher = {American Institute of Physics},
  doi = {10.1063/1.523763}
}

@article{Page1993,
  title = {Average entropy of a subsystem},
  author = {Page, Don N.},
  journal = {Physical Review Letters},
  volume = {71},
  number = {9},
  pages = {1291--1294},
  year = {1993},
  publisher = {American Physical Society},
  doi = {10.1103/PhysRevLett.71.1291}
}

@article{Horodecki2009,
  title   = {Quantum entanglement},
  author  = {Horodecki, Ryszard and Horodecki, Pawe{\l} and Horodecki, Micha{\l} and Horodecki, Karol},
  journal = {Reviews of Modern Physics},
  volume  = {81},
  number  = {2},
  pages   = {865--942},
  year    = {2009}
}

@article{Plenio2007,
  title   = {An introduction to entanglement measures},
  author  = {Plenio, Martin B. and Virmani, S.},
  journal = {Quantum Information \& Computation},
  volume  = {7},
  number  = {1},
  pages   = {1--51},
  year    = {2007}
}

@article{Giovannetti2004,
  title   = {Quantum-enhanced measurements: beating the standard quantum limit},
  author  = {Giovannetti, Vittorio and Lloyd, Seth and Maccone, Lorenzo},
  journal = {Science},
  volume  = {306},
  number  = {5700},
  pages   = {1330--1336},
  year    = {2004}
}

@article{Wineland1992,
  title   = {Spin squeezing and reduced quantum noise in spectroscopy},
  author  = {Wineland, D. J. and Bollinger, J. J. and Itano, W. M. and Moore, F. L. and Heinzen, D. J.},
  journal = {Physical Review A},
  volume  = {46},
  number  = {11},
  pages   = {R6797--R6800},
  year    = {1992}
}

@article{Pezze2018,
  title   = {Quantum metrology with nonclassical states of atomic ensembles},
  author  = {Pezz{\`e}, Luca and Smerzi, Augusto and Oberthaler, Markus K. and Schmied, Roman and Treutlein, Philipp},
  journal = {Reviews of Modern Physics},
  volume  = {90},
  number  = {3},
  pages   = {035005},
  year    = {2018}
}

@article{Ma2011,
  title   = {Quantum spin squeezing},
  author  = {Ma, Jian and Wang, Xiaoguang and Sun, C. P. and Nori, Franco},
  journal = {Physics Reports},
  volume  = {509},
  number  = {2-3},
  pages   = {89--165},
  year    = {2011}
}

@article{Gross2010,
  title   = {Nonlinear atom interferometer surpasses classical precision limit},
  author  = {Gross, Christian and Zibold, Tilman and Nicklas, Eike and Est{\`e}ve, J{\'e}r{\^o}me and Oberthaler, Markus K.},
  journal = {Nature},
  volume  = {464},
  number  = {7292},
  pages   = {1165--1169},
  year    = {2010}
}

@article{Riedel2010,
  title   = {Atom-chip-based generation of entanglement for quantum metrology},
  author  = {Riedel, Max F. and others},
  journal = {Nature},
  volume  = {464},
  number  = {7292},
  pages   = {1170--1173},
  year    = {2010}
}

@article{Popescu2006,
  title   = {Entanglement and the foundations of statistical mechanics},
  author  = {Popescu, Sandu and Short, Anthony J. and Winter, Andreas},
  journal = {Nature Physics},
  volume  = {2},
  number  = {11},
  pages   = {754--758},
  year    = {2006}
}

@article{Gnatenko26,
title = {Studies of properties of bipartite graphs with quantum programming},
journal = {Physics Letters A},
volume = {566},
pages = {131191},
year = {2026},
issn = {0375-9601},
doi = {https://doi.org/10.1016/j.physleta.2025.131191},
author = {Kh. P. Gnatenko}
}

@article{Gnatenko24,
title = {Entanglement of multi-qubit states representing directed networks and its detection with quantum computing},
journal = {Physics Letters A},
volume = {521},
pages = {129815},
year = {2024},
issn = {0375-9601},
doi = {https://doi.org/10.1016/j.physleta.2024.129815},
author = {Kh.P. Gnatenko}
}

@misc{Gnatenko23,
    author = "Gnatenko, Kh. P.",
    title = "{Evaluation of variational quantum states entanglement on a quantum computer by the mean value of spin}",
    eprint = "2301.03885",
    archivePrefix = "arXiv",
    primaryClass = "quant-ph",
    month = "1",
    year = "2023"
}

@article{GnatenkoL22,
title = {Geometric properties of evolutionary graph states and their detection on a quantum computer},
journal = {Physics Letters A},
volume = {452},
pages = {128434},
year = {2022},
issn = {0375-9601},
doi = {https://doi.org/10.1016/j.physleta.2022.128434},
author = {Kh.P. Gnatenko and H.P. Laba and V.M. Tkachuk}
}

@article{Gnatenko21,
doi = {10.1209/0295-5075/ac419b},
year = {2022},
month = {mar},
publisher = {EDP Sciences, IOP Publishing and Società Italiana di Fisica},
volume = {136},
number = {4},
pages = {40003},
author = {Gnatenko, Kh. P. and Susulovska, N. A.},
title = {Geometric measure of entanglement of multi-qubit graph states and its detection on a quantum computer},
journal = {Europhysics Letters}
}

@article{DeSimone2025,
author = {De Simone, Lucio and Franzosi, Roberto},
title = {Entanglement in Quantum Systems Based on Directed Graphs},
journal = {Advanced Quantum Technologies},
year = 2025,
pages = {e00514},
doi = {https://doi.org/10.1002/qute.202500514}
}

@Article{DeSimone2026,
AUTHOR = {De Simone, Lucio and Capra, Lorenzo and Vesperini, Arthur and Rossi, Leonardo and Di Cairano, Loris and Franzosi, Roberto},
TITLE = {Geometric Aspects of Entanglement},
JOURNAL = {Entropy},
VOLUME = {28},
YEAR = {2026},
NUMBER = {3},
ARTICLE-NUMBER = {299},
URL = {https://www.mdpi.com/1099-4300/28/3/299},
ISSN = {1099-4300},
DOI = {10.3390/e28030299}
}

@article{Vesperini2024b,
author = {Vesperini, Arthur and Franzosi, Roberto},
title = {Entanglement, Quantum Correlators, and Connectivity in Graph States},
journal = {Advanced Quantum Technologies},
volume = {7},
number = {2},
pages = {2300264},
year = {2024}
}

@misc{Gori2024,
      title={Gaussian Entanglement Measure: Applications to Multipartite Entanglement of Graph States and Bosonic Field Theory}, 
      author={Matteo Gori and Matthieu Sarkis and Alexandre Tkatchenko},
      year={2024},
      eprint={2401.17938},
      archivePrefix={arXiv},
      primaryClass={quant-ph},
      url={https://arxiv.org/abs/2401.17938}, 
}

@article{berry,
    author = {Berry, Michael Victor},
    title = {Quantal phase factors accompanying adiabatic changes},
    journal = {Proceedings of the Royal Society of London. A. Mathematical and Physical Sciences},
    volume = {392},
    number = {1802},
    pages = {45-57},
    year = {1984},
    month = {03},
    issn = {0080-4630},
    doi = {10.1098/rspa.1984.0023},
    url = {https://doi.org/10.1098/rspa.1984.0023},
    eprint = {https://royalsocietypublishing.org/rspa/article-pdf/392/1802/45/65511/rspa.1984.0023.pdf},
}

@article{PhysRev.115.485,
  title = {Significance of Electromagnetic Potentials in the Quantum Theory},
  author = {Aharonov, Y. and Bohm, D.},
  journal = {Phys. Rev.},
  volume = {115},
  issue = {3},
  pages = {485--491},
  numpages = {0},
  year = {1959},
  month = {Aug},
  publisher = {American Physical Society},
  doi = {10.1103/PhysRev.115.485},
  url = {https://link.aps.org/doi/10.1103/PhysRev.115.485}
}

@article{PhysRevLett.49.405,
  title = {Quantized Hall Conductance in a Two-Dimensional Periodic Potential},
  author = {Thouless, D. J. and Kohmoto, M. and Nightingale, M. P. and den Nijs, M.},
  journal = {Phys. Rev. Lett.},
  volume = {49},
  issue = {6},
  pages = {405--408},
  numpages = {0},
  year = {1982},
  month = {Aug},
  publisher = {American Physical Society},
  doi = {10.1103/PhysRevLett.49.405},
  url = {https://link.aps.org/doi/10.1103/PhysRevLett.49.405}
}

@article{ZANARDI199994,
title = {Holonomic quantum computation},
journal = {Physics Letters A},
volume = {264},
number = {2},
pages = {94-99},
year = {1999},
issn = {0375-9601},
doi = {https://doi.org/10.1016/S0375-9601(99)00803-8},
url = {https://www.sciencedirect.com/science/article/pii/S0375960199008038},
author = {Paolo Zanardi and Mario Rasetti}
}

@article{gibbons,
title = "Typical states and density matrices",
journal = "Journal of Geometry and Physics",
volume = "8",
number = "1",
pages = "147 - 162",
year = "1992",
issn = "0393-0440",
doi = "10.1016/0393-0440(92)90046-4",
url = "http://www.sciencedirect.com/science/article/pii/0393044092900464",
author = "G.W. Gibbons"
}

\end{document}